\documentclass[namedreferences]{SolarPhysics}
\usepackage[optionalrh]{spr-sola-addons} % For Solar Physics 
\usepackage{epsfig}                     % For eps figures, old commands
\usepackage{graphicx}                    % For eps figures, newer & more powerfull
\usepackage{color}                       % For color text: \color command
\usepackage{url}                         % For breaking URLs easily trough lines
\usepackage{rotating}
\begin{document}

\begin{article}

\begin{opening}

\title{Heliospheric Observations of STEREO-Directed Coronal Mass Ejections in 2008--2010: Lessons for Future Observations of Earth-Directed CMEs}

%%%%%%%%%%%%%%%%%%%%%%%%%%%%%%%%%%%%%%%%%%%%%%%%%%%
%% Authors Names
%
\author{N.~\surname{Lugaz}$^{1,2}$\sep P.~\surname{Kintner}$^{2,3}$\sep C.~\surname{M{\"o}stl}$^{4,5,6}$\sep L.~K.~\surname{Jian}$^{7,8}$\sep  C.~J.~\surname{Davis}$^{9,10}$\sep C.~J.~\surname{Farrugia}$^{1}$}

%%%%%%%%%%%%%%%%%%%%%%%%%%%%%%%%%%%%%%%%%%%%%%%%%%%
%% Runningheads
%
\runningauthor{Lugaz et al.}
\runningtitle{Heliospheric Observations of STEREO Events}

%%%%%%%%%%%%%%%%%%%%%%%%%%%%%%%%%%%%%%%%%%%%%%%%%%%
%% Affilations 
%
\institute{$^{1}$ Space Science Center, University of New Hampshire, Durham, New Hampshire, USA
                     email: \url{noe.lugaz@unh.edu}; \url{charlie.farrugia@unh.edu}\\                
                     $^{2}$ Institute for Astronomy, University of Hawaii, 2680 Woodlawn Dr., Honolulu, HI 96822, USA\\
                     $^{3}$ University of Rochester, Rochester, NY 14627, USA
                     email: \url{pkintner@u.rochester.edu}\\
                     $^{4}$ Space Science Laboratory - University of California, Berkeley, 94720 CA, USA
                     email: \url{christian.moestl@ssl.berkeley.edu}\\
                     $^{5}$Kanzelh\"ohe Observatory-IGAM, Institute of Physics, University of Graz, Universit\"atsplatz 5, A-8010, Graz, Austria\\
                     $^{6}$Space Research Institute, Austrian Academy of Sciences, Graz 8042, Austria\\
                     $^{7}$ Department of Astronomy, University of Maryland, College Park, Maryland, USA
                     email: \url{lan.jian@nasa.gov}\\
                     $^{8}$  Heliophysics Science Division, NASA Goddard Space Flight Center, Greenbelt, Maryland, USA\\
		  $^{9}$ SFTC Rutherford Appleton Laboratory, Didcot, Oxfordshire OX11 0QX, UK
                       email: \url{chris.davis@stfc.ac.uk}\\  
                     $^{10}$ Department of Meteorology, University of Reading, Berkshire, RG6 7BE}                     
                     
                     %$^{7}$ University of Maryland, College Park, MD\\

%%%%%%%%%%%%%%%%%%%%%%%%%%%%%%%%%%%%%%%%%%%%%%%%%%%
%%% Abstract 
\begin{abstract}

We present a study of coronal mass ejections (CMEs) which impacted one of the STEREO spacecraft between January 2008 and early 2010. We focus our study on 20 CMEs which were observed remotely by the {\it Heliospheric Imagers} (HIs) onboard the other STEREO spacecraft up to large heliocentric distances. We compare the predictions of the Fixed-$\Phi$ and Harmonic Mean (HM) fitting methods, which only differ by the assumed geometry of the CME. It is possible to use these techniques to determine from remote-sensing observations the CME direction of propagation, arrival time and final speed which are compared to {\it in situ} measurements. 
We find evidence that for large viewing angles, the HM fitting method predicts the CME direction better. However, this may be due to the fact that only wide CMEs can be successfully observed when the CME propagates more than 100$^\circ$ from the observing spacecraft.
Overall eight CMEs, originating from behind the limb as seen by one of the STEREO spacecraft can be tracked and their arrival time at the other STEREO spacecraft can be successfully predicted. This includes CMEs, such as the events on 4 December 2009 and 9 April 2010,  which were viewed 130$^\circ$ away from their direction of propagation. Therefore, we predict that some Earth-directed CMEs will be observed by the HIs until early 2013, when the separation between Earth and one of the STEREO spacecraft will be similar to the separation of the two STEREO spacecraft in 2009--2010. 

\end{abstract}
\end{opening}

%% ------------------------------------------------------------------------ %%
%
%  BEGIN ARTICLE
%
%% ------------------------------------------------------------------------ %%

\section{Introduction} \label{intro}

Remote observations of coronal mass ejections (CMEs) by heliospheric imagers (HIs), combined with {\it in situ} measurements provide a unique opportunity to study the evolution, propagation and properties of these transients. It also constitutes a main science objective of the {\it Solar-Terrestrial Relations Observatory} (STEREO) mission. White-light observations by the HIs onboard the STEREO  and by the {\it Solar Mass Ejection Imager} (SMEI) onboard {\it Coriolis} can be analyzed to determine the direction of propagation of CMEs and their kinematics (e.g. with the methods of \opencite{Rouillard:2008}, \opencite{THoward:2009b},  \opencite{Maloney:2009}, \opencite{Liu:2010b}, \opencite{Lugaz:2010b}, \opencite{Lugaz:2010c}). These methods can be used in real-time with beacon data from STEREO to predict whether or not a CME will hit a spacecraft, and, if a hit is forecasted, to predict the arrival time and speed at 1~AU \cite{Davis:2011}. {\it In situ} measurements can then be used to test, compare and validate these predictions. Another approach is to constrain remote-sensing observations with the knowledge obtained from {\it in situ} measurements (final speed, arrival time). This is the approach taken, for example, in \inlinecite{Wood:2009a}, \inlinecite{Moestl:2010},  \inlinecite{Temmer:2011} and \inlinecite{Rollett:2012}  to study CME kinematics. 

Such coordinated studies have been primarily undertaken for Earth-directed CMEs, which impacted ACE or {\it Wind} \cite{Davis:2009,Rouillard:2009,Wood:2009a,Moestl:2010,Liu:2010b,Liu:2011} but also for CMEs or corotating interaction regions (CIRs) which impacted {\it Venus Express} \cite{Rouillard:2009b,Moestl:2011}, MESSENGER or spacecraft orbiting the planet Mars \cite{Williams:2011}. Most of the analyses of data from planetary missions have been for CMEs which also impacted Earth. There have also been a few studies of CMEs which impacted one of the STEREO spacecraft \cite{Wood:2009b,Moestl:2009b}. %In the next few years, with the launch of Mars Atmosphere and Volatile EvolutioN (MAVEN) \cite{Jakosky:2009}, there should be additional studies of Mars-directed CMEs. 

As the separation between the two STEREO spacecraft increases, stereoscopic views of a CME into the HI field-of-view are expected to become rarer. Therefore, stereoscopic methods \cite{Liu:2010b,Lugaz:2010b} making use of the two STEREO views may not be applicable for Earth-directed CMEs from mid-2011 onwards, when the total separation is larger than 180$^\circ$. In addition, it is likely that future missions (e.g., see \opencite{Gopalswamy:2011}) will only have one spacecraft with a HI instrument. It is therefore important to further test and validate methods using HI observations from a single spacecraft. The two main fitting methods, the Fixed-$\Phi$ (F$\Phi$; \opencite{Rouillard:2008}), and Harmonic Mean (HM; \opencite{Lugaz:2010c}), have been compared for three Earth-directed CMEs in \inlinecite{Rollett:2012} and for CMEs observed simultaneously by both STEREO spacecraft in 2008 and 2009 \cite{Lugaz:2010c}. It was found that  the two methods differ for small ($< 30^\circ$) and large ($>80^\circ$) viewing angles \cite{Lugaz:2010c,Davies:2012}. The largest viewing angle for which these two techniques have been compared for a Earth-impacting CME is about 67$^\circ$ for the April 2010 CME. This spacecraft separation is within the range where the techniques are expected to give similar results. Studying STEREO-directed events has the advantage of allowing for larger viewing angles compared to Earth-directed CMEs and to prepare for future observations of Earth-directed CMEs in future years. %The first quadrature observations was obtained between the two STEREO spacecraft in June 2009 and paved the way for the Earth-STEREO quadrature in January 2011. 

Here, we analyze the remote-sensing observations of STEREO-impacting CMEs during the years 2008--2010. We determine the CME direction of propagation and speed by fitting the remote-sensing observations with the two fitting techniques (F$\Phi$ and HM). The predicted hit/miss, arrival time and speed are then compared with {\it in situ} measurements. In this way, we are able: i) to assess which of the two methods is able to better determine some of the CME properties measured {\it in situ} and, ii) to determine how the accuracy of the methods is affected by the increasing separation of the STEREO spacecraft from the Sun-Earth line. We use STEREO-directed events from 2009 to 2010 to prepare for the large separation between the Earth and the STEREO spacecraft which will reach 125$^\circ$ by January 2013. This allows us to establish what might be the largest viewing angle for which a CME can be observed into HI-2.

In Section \ref{sec:data}, we summarize briefly the two fitting techniques used in this article and explain our data selection process. In Section \ref{sec:Dec2009}, we present the detailed analysis of one event, which occurred on 4 December 2009, when the  separation between the two STEREO spacecraft was 130$^\circ$. In Section \ref{sec:stat}, we present the results from the analysis of 20 events which occurred between the beginning of 2008 and mid-2010. The conclusions of our investigation are drawn in Section \ref{sec:conclusions}.

%%%%%%%%%%%%%%%%%%%%%%%%%%%%%%%%%%%%%%%%%%%%%%%%%%%%%%%%%%%%%%%%%%%%%%%%%%%%
\section{Analysis Techniques and Data Selection} \label{sec:data}
%%%%%%%%%%%%%%%%%%%%%%%%%%%%%%%%%%%%%%%%%%%%%%%%%%%%%%%%%%%%%%%%%%%%%%%%%%%%
\subsection{Fitting Methods}

In the field-of-view of a heliospheric imager, the position of a density feature (e.g., a Stream Interaction Region --SIR-- or a CME) is measured as the angle between the observing spacecraft, the Sun and the density structure, and it is commonly referred to as the elongation angle. When CMEs are observed to large elongation angles, the time-elongation data can be fitted to analytical functions and a single value for the CME speed and direction of propagation can be derived \cite{Sheeley:1999}. Two common methods are the F$\Phi$ \cite{Rouillard:2008} and the HM fittings \cite{Lugaz:2010c}. We will focus here on these two methods only. A schematic view of the two methods is shown in the bottom panel of Figure~1. Here, we use the following notation: $\alpha$ is the elongation angle, $\Phi$  the direction of propagation,  which is assumed to be fixed, $d_\mathrm{ST}$ the heliocentric position of the observing spacecraft (STEREO) and $t$ the time.
 
Assuming a constant propagation speed, $V$, and a single plasma element, the relation between elongation and distance of \inlinecite{Kahler:2005} can be inverted to obtain the F$\Phi$ fitting relation \cite{Rouillard:2008}:

\begin{equation}
%\alpha = \arctan \left (\frac{ V_{\mathrm{F}\Phi F}  t \sin\Phi_{\mathrm{F}\Phi F} }{d_\mathrm{ST} - V_{\mathrm{F}\Phi F} t \cos\Phi_{\mathrm{F}\Phi F}} \right).\label{eq:TEFF}
\alpha = \arctan \left (\frac{ V  t \sin\Phi }{d_\mathrm{ST} - V t \cos\Phi} \right).\label{eq:TEFF}
\end{equation}
 
A measured time-elongation profile can be fitted to a profile of calculated elongations given by Equation~(1). The F$\Phi$ approximation assumes that the CME is a single point and therefore, it is not straightforward to determine when the fitting results implies a hit or a miss and what is the expected arrival time and speed of different parts of the CME. Previous works \cite{Davies:2009,Moestl:2011} made the assumption that the F$\Phi$ fitting predicts a hit if the best-fitted direction is within $10^\circ$--$15^\circ$ of a spacecraft and that the arrival time and speed are the same for all parts of the ``front''.  Here, we use a looser criterion of $\pm 20^\circ$ around the best-fit direction and we also assume that all parts of the CME ``front'' arrive at the same time with the same speed. This criterion of $\pm 20^\circ$ is used to include the uncertainty in the best-fit direction of propagation, which is typically 10--15$^\circ$ and the fact that even narrow density features are expected to have a half-width of  about 10$^\circ$. Assuming a half-width wider than 20$^\circ$ is not consistent with the underlying assumption of the F$\Phi$ approximation: the fact that the observed signal originates from the apex of the CME. This can be seen in the bottom panel of Figure~1 for a large viewing angle: if the CME is assumed to be wide, the wings of the CME are at a larger elongation angle than the apex of the CME (see also discussion in \opencite{THoward:2009a}). 
%It should be noted that, for CMEs observed at large viewing angles, assuming that the distance is given by the Fixed-$\Phi$ approximation and that the CME front extends $\pm 20^\circ$ from the CME apex is not self-consistent, but there are no simple alternatives to these two assumptions.
 
Another simple assumption for what is observed by HIs, as proposed by \inlinecite{THoward:2009b} and \inlinecite{Lugaz:2009c}, is as follows. The CME front can be modeled as a locally circular front  with a diameter equal to the distance of the apex (or equivalently, a front which is anchored at the Sun). It is further assumed that the measured elongation angle simply corresponds to the angle between the Sun-spacecraft line and {\em the line-of-sight tangent to this circular front}. 
Assuming a constant propagation speed and the geometry explained above, the HM fitting relation can be obtained \cite{Lugaz:2010c,Liu:2010b,Moestl:2011}:

\begin{eqnarray}
%\alpha &= & \arctan \left (\frac{ V_\mathrm{HMCV}  t \sin\beta_\mathrm{HMCV} }{2 d_\mathrm{ST} - V_\mathrm{HMCV} t \cos\beta_\mathrm{HMCV}} \right) +  \label{eq:TEHM} \\
%& & \arcsin \left (\frac{ V_\mathrm{HMCV} t  }{\sqrt{\left(2 d_\mathrm{ST} - V_\mathrm{HMCV} t \cos\beta_\mathrm{HMCV} \right)^2 + \left(V_\mathrm{HMCV}  t \sin\beta_\mathrm{HMCV} \right)^2}} \right). \nonumber
\alpha &= & \arctan \left (\frac{ V  t \sin\Phi}{2 d_\mathrm{ST} - V t \cos\Phi} \right) +  \label{eq:TEHM} \\
& & \arcsin \left (\frac{ V t  }{\sqrt{\left(2 d_\mathrm{ST} - V t \cos\Phi \right)^2 + \left(V  t \sin\Phi \right)^2}} \right). \nonumber
\end{eqnarray}

A measured time-elongation profile can also be fitted to profiles of calculated elongations  given by Equation~(2). The fitted speed, $V_\mathrm{best-fit}$, and direction, $\Phi_\mathrm{best-fit}$, correspond to that of the nose (apex) of the CME. To determine the predicted speed and arrival time at a spacecraft which is not hit directly by the nose of the CME, a correction is needed as derived in \inlinecite{Moestl:2011}. For a spacecraft at a heliocentric distance $d_\mathrm{spacecraft}$ and separated by an angle $\beta$ with the observing spacecraft, the predicted speed is $V_\mathrm{spacecraft} = V_\mathrm{best-fit} \cos(\beta - \Phi_\mathrm{best-fit} )$ and the predicted time is $t_\mathrm{arr} = d_\mathrm{spacecraft}/V_\mathrm{spacecraft}$.

It should be noted that both fitting methods assume a constant propagation speed and a constant direction. Using stereoscopic methods \cite{Liu:2010b,Lugaz:2010b,Byrne:2010}, it is possible to relax these two assumptions and derive for all pairs of observations the CME direction and position (and speed). However, it is unlikely that the two STEREO spacecraft can observe a CME into HI-2 field-of-view when the separation between the spacecraft is 180$^\circ$ or beyond. The technique of  \inlinecite{THoward:2009a} can be used to derive the CME position without assuming a constant propagation speed, although it is not straightforward to decouple the effects of the varying speed from the change in the CME shape. Due to their simplicity, fitting techniques are used for real-time space weather forecasting \cite{Davis:2011} and for scientific analyses by a number of groups \cite{Wood:2009b,Temmer:2011,Rollett:2012} and we focus here on these techniques. 

%%%%%Figure 1%%%%%%%%%%%%%%%%%%%
\begin{figure*}[t*]
\begin{center}
{\includegraphics*[width=7cm]{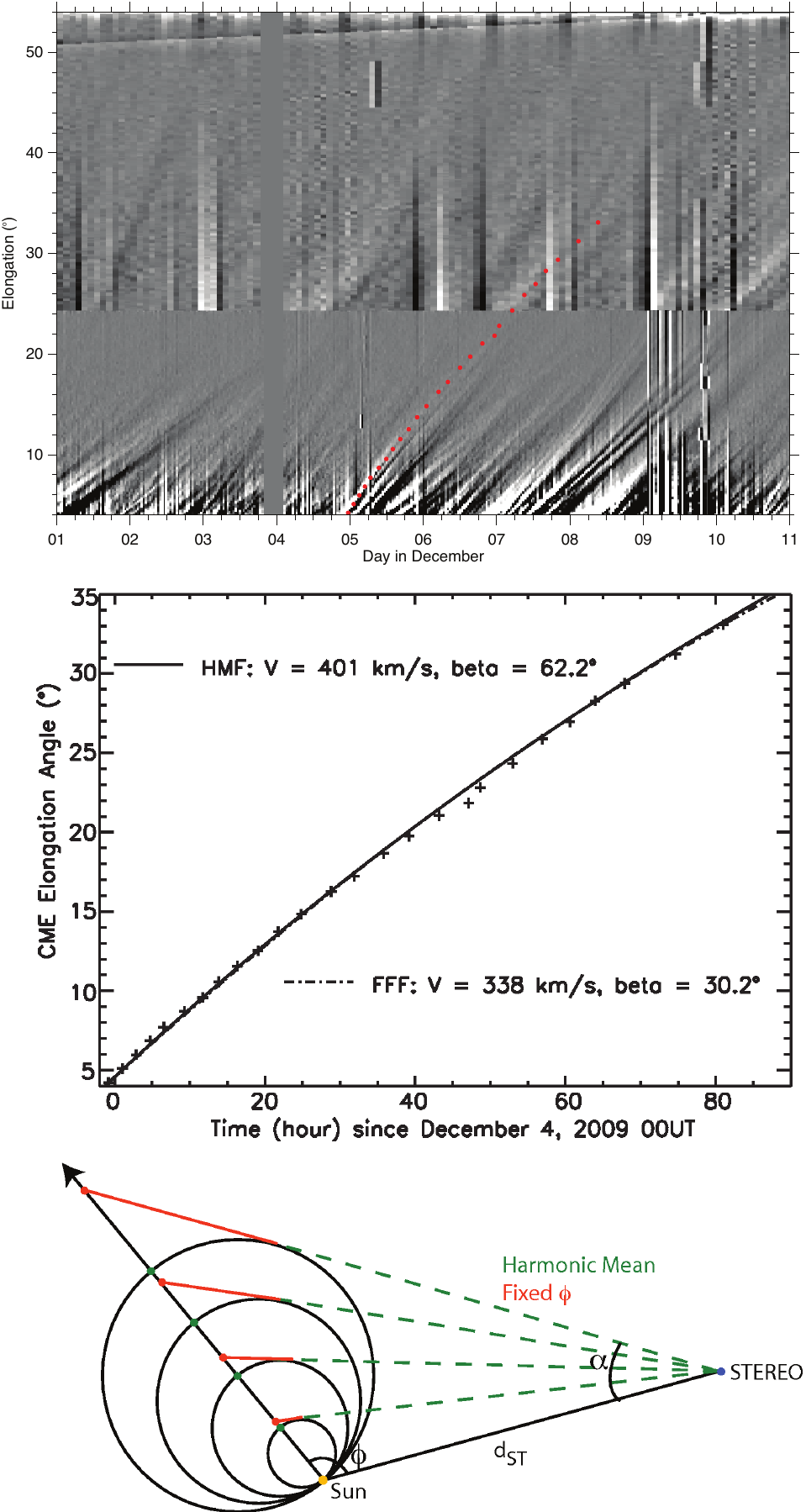}}
\caption{Top and Middle: Analysis of the 4 December 2009 CME observed by STEREO-B. Top: J-map and datapoints (red circle) corresponding to the ``black'' front. Middle: Datapoints (plus symbols) and best-fit solution obtained with HM fitting method (solid line) and F$\Phi$ fitting method (dash-dot line) fittings. In this panel, beta refers to the best-fit direction with respect to the Sun-Earth line. Bottom: Sketch of the Fixed-$\Phi$ (orange) and Harmonic Mean (green) approximations.}
\end{center}
\end{figure*}
%%%%%%%%%%%%%%%%%%%%%%%%%

\subsection{Data Selection}

Following other researchers (e.g., see \opencite{Richardson:2010}), we start from ICMEs measured {\it in situ} at 1~AU by one of the two STEREO spacecraft and attempt to identify, first, their heliospheric sources as measured in the HI instruments and, second, their coronal sources. Our starting point is the list of ICMEs observed by IMPACT \cite{Luhmann:2008} and PLASTIC \cite{Galvin:2008} onboard STEREO as maintained at \url{www-ssc.igpp.ucla.edu/~jlan/STEREO/Level3}. The determination of the ICMEs and their characteristics can be found in \inlinecite{Jian:2006}. We focus on ICMEs detected from January 2008 to July 2010: 21 by STEREO-A and 26 by STEREO-B. For this list of 47 ICMEs, we attempt to find their heliospheric counterparts in remote-sensing images. We use the list summarized on the Rutherford Appleton Laboratory (RAL) website \url{www.stereo.rl.ac.uk/HIEventList.html } of heliospheric transients observed by the HIs \cite{Eyles:2009} part of SECCHI \cite{Howard:2008} onboard STEREO. This list consists of the time-elongation tracks measured on J-maps \cite{Davies:2009} along elevation 0$^\circ$ for various transients. Some of the time-elongation measurements from this list have been used in previous studies \cite{Davis:2010,Lugaz:2010c}. On the RAL website, the time-elongation datapoints for each track are plotted; in the present study, we use the corresponding time-elongation measurements.

J-maps are constructed using running differences of SECCHI images; intensity enhancements with respect to the previous image appear as bright regions while depletions appear as black regions \cite{Davies:2009}. As noted in \inlinecite{Davis:2010}, the clearest feature to track is the ``black'' front, i.e. the transition between the bright regions and the dark regions (maximum contrast). This ``black'' front is typically 4--6 hours behind the start of the CME and, therefore, our tracking introduces a systematic offset in the predicted arrival time of an event. A J-map and an example of the tracking of a ``black'' front (with red circles) are shown in the top panel of Figure~1. In \inlinecite{Lugaz:2010b}, we compared the direction of propagation and distance derived by stereoscopic methods for the 2008 April 26 CME (CME \# 2 in the present study) using the bright front and the ``black'' front of the CME. We found that in the HI fields-of-view the kinematics and direction of the CME using the two tracks are nearly identical. As a summary, in order to follow the CMEs to large elongation angles, we use the tracks of the ``black'' front of the CMEs; this introduces a systematic offset in the predicted arrival time but it is not expected to change the predicted arrival speed and CME direction. Since no study has yet focused on determining the exact value for this offset, we do not correct the predicted arrival times for it. 

For each ICME from the IMPACT list, we have looked in the RAL list for transients whose starting time is within a window of 36 to 120 hours prior to the arrival time of the ICME. We use HI-1 movies to identify the CMEs and to associate them with one (or multiple) of the tracks from the RAL list.  %The RAL website lists the best-fit direction for the F$\Phi$ fitting method. The best-fit direction for the HM fitting method can be estimated using the F$\Phi$ direction, for example using Equation~(4) from \inlinecite{Lugaz:2010c}. These two directions are used to select the RAL track most likely to be associated with an ICME from the IMPACT list. 
In most cases, there is only one CME for which the expected arrival time is within $\pm 24$~hours  of the actual  arrival time and whose time-elongation profile shows the required trend (deceleration for large viewing angles, acceleration for small ones). However, there are often multiple tracks associated with each CME, corresponding, for example, to the two dense fronts which bracket the CMEs (see, e.g., \opencite{Savani:2011a}) and sometimes to core material. In these cases, we have chosen the track corresponding to the leading edge of the CME. Sometimes, there are no associated tracks, either because no CME was visible in the field-of-view of the HIs or because there is no track in the RAL list associated with a given CME (often because the CME is faint and not visible to large elongation angles, sometimes because of data gaps). Overall, we find 20 ICMEs ($\sim 40\%$) for which there is a corresponding track in the RAL list. Using information from the HI movies, we identify the CMEs in COR-2 and COR-1, when possible. The 20 CMEs as well as their first appearance in COR-1 or COR-2 are summarized in Table~1.  Some of the CMEs are streamer blowouts, which typically form at heights of 2--3~R$_\odot$, start at low speed and may not be associated with any clear signature in the low corona (e.g., see \opencite{Robbrecht:2009}). For these CMEs, we do not indicate a starting time but only a starting day in Table~1. The information from COR-1 and COR-2 is not directly used in this study, but it is listed for completeness and, for future references, if our list is to be used in future studies. In Table~ 1, CMEs are ordered by the total separation between the two STEREO spacecraft. Since each CME was observed remotely by one STEREO spacecraft and measured {\it in situ} by the other, the total separation gives an estimate of the viewing angle at which STEREO remotely observed the CME.

The proportion of ICMEs for which we find remote-sensing observations  into the outer heliosphere allowing a successful prediction decreases with increasing separation between the STEREO spacecraft, from 10/14 in 2008 to 9/21 in 2009 and only 1/12 in 2010. However, there are 5 CMEs which were tracked into HI-2 field-of-view with a spacecraft separation of more than 120$^\circ$. %We discuss one CME event for which we find the solar and heliospheric sources but the fitting methods failed in section \ref{sec:later}.

\begin{table}[t]
\begin{tabular}{cccccccc}
\hline
CME &  IMPACT  & RAL & 1$^\mathrm{st}$ Detection & Tot. Separation\\
\hline
{\bf 2008}\\
1 & 7B & 327A & 31 Jan.  (COR-2B) & 45.3$^\circ$\\
2 & 9B & 399A & 26 Apr.  14:25 (COR-1B)& 49.8$^\circ$\\
3 & 2A & 172B & 7 May  (COR-2A) & 51$^\circ$\\
4 & 11B & 428A & 1 Jun. 10:25 (COR-1B) & 54.8$^\circ$\\
5 & 3A & 214B &  30 Jun.  (COR-2A) &58.3$^\circ$ \\
6 & 12B & 514A & 10 Aug.  (COR-2B) & 67.2$^\circ$\\
7 & 4A & 247B & 30 Aug.  16:05 (COR-1A) & 71.3$^\circ$\\
8 & 5A & 270B & 26 Oct.  (COR-1A)  & 81.3$^\circ$\\
9 & 6A & 290B & 23 Nov.  11:25 (COR-1A) & 85.2$^\circ$\\
10 & 14B & 595A & 27 Dec.  02:25 (COR-1B) & 88.4$^\circ$\\
\hline
{\bf 2009}\\
11 & 10B & 611A & 8 Jan. $\sim$ 4:00 (COR-1B) & 89.3$^\circ$\\
12 & 1A & 328B & 21 Jan. 16:45 (COR-1A, f?) & 90$^\circ$\\
13 & 12B & 762A & 10 Jul. (COR-2B) & 104.1$^\circ$\\
14 & 13B & 771A & 26 Jul. 11:35 (COR-2B) & 106.5$^\circ$\\
15 & 18B & 818A & 26 Sep. 17:45 (COR-1B, f) & 118.4$^\circ$\\
16 & 19B & 858A & 5 Nov.  8:05 (COR-1B, f) & 125.3$^\circ$\\
17 & 7A & 466B & 8 Nov.  5:25 (COR-1A,f) & 125.8$^\circ$\\
18 & 20B & 881A & 22 Nov. (COR-2B) & 127.9$^\circ$\\
19 & 9A & 489B & 4 Dec.  6:50 (COR-1A, f)& 129.5$^\circ$\\
\hline
{\bf 2010}\\ 
20 & 3A & 575B & 19 Apr. 14:25 (COR-1A,f)  & 139.6$^\circ$\\
\hline
\end{tabular}
\caption{\underline{STEREO-impacting and STEREO-observed CME list:} The columns show the CME number in this study, the ICME number (with year) from the IMPACT list, the track number from the RAL list, the date of first detection in COR-1 or COR-2 and the separation between STEREO-A and STEREO-B from left to right. (f) indicates a filament eruption.}
\end{table}

%%%%%Figure 1%%%%%%%%%%%%%%%%%%%
\begin{figure*}[t*]
\begin{center}
{\includegraphics*[width=6cm]{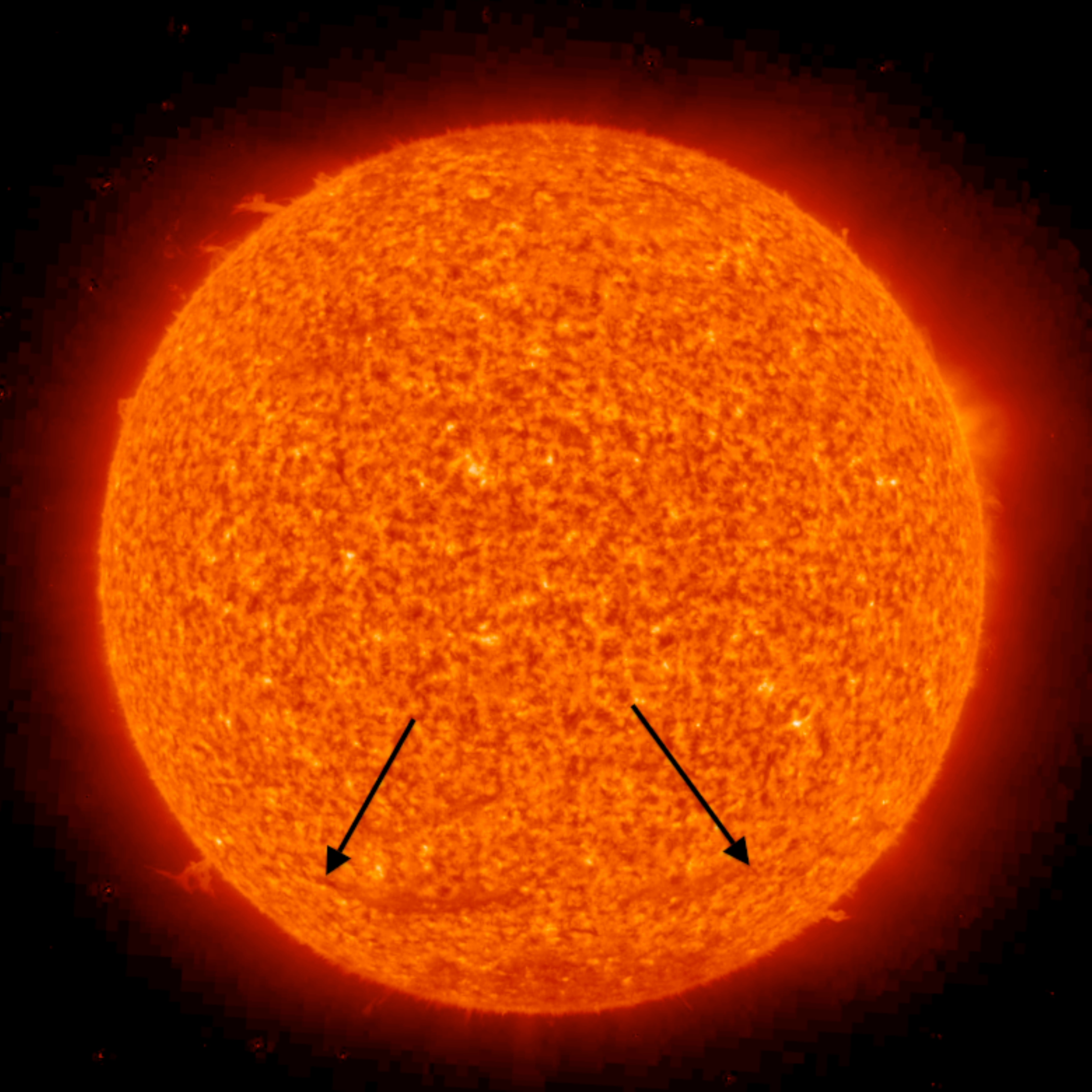}}
{\includegraphics*[width=6cm]{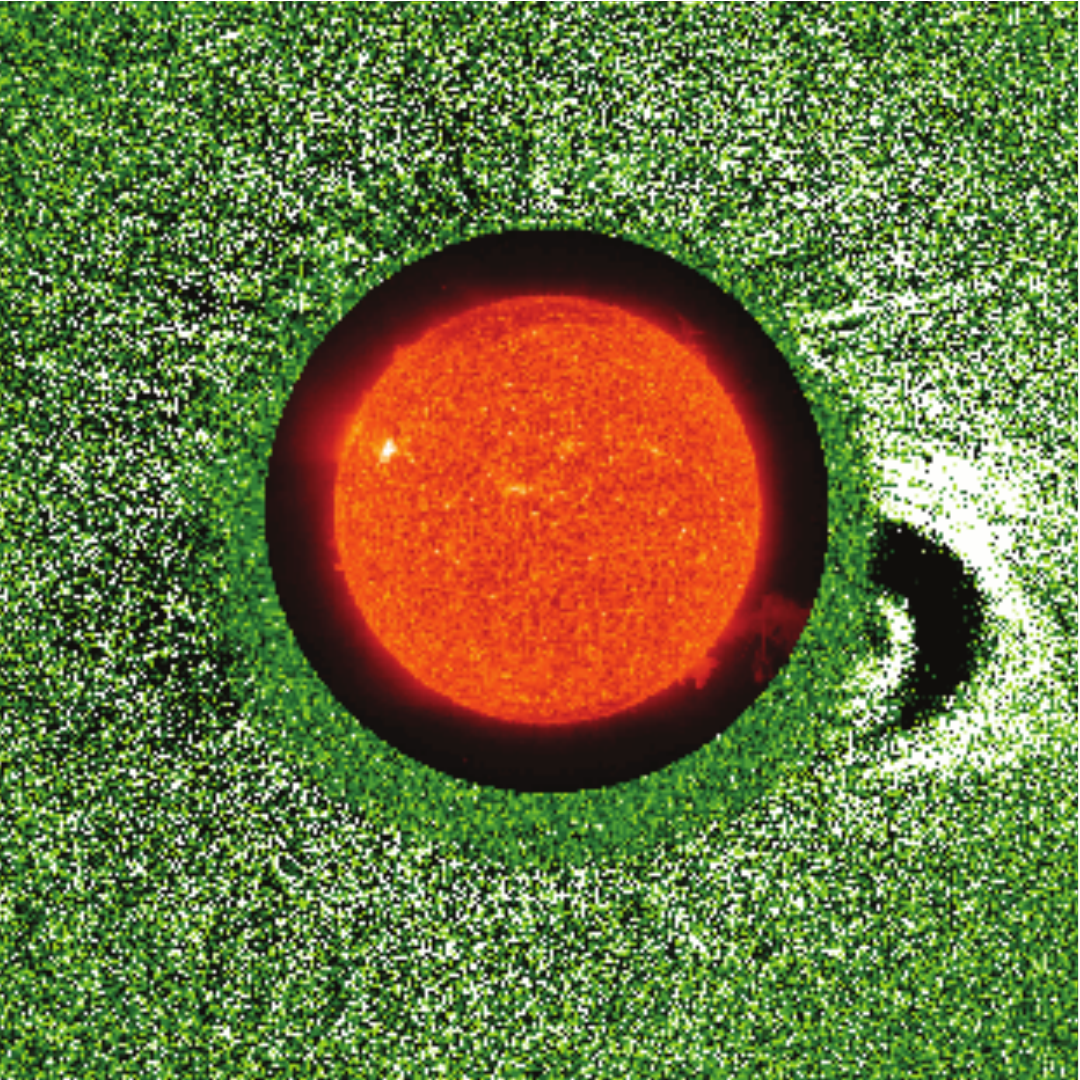}}\\
{\includegraphics*[width=6cm]{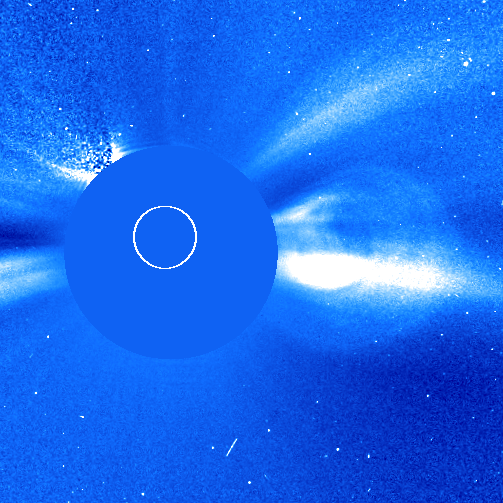}}
{\includegraphics*[width=6cm]{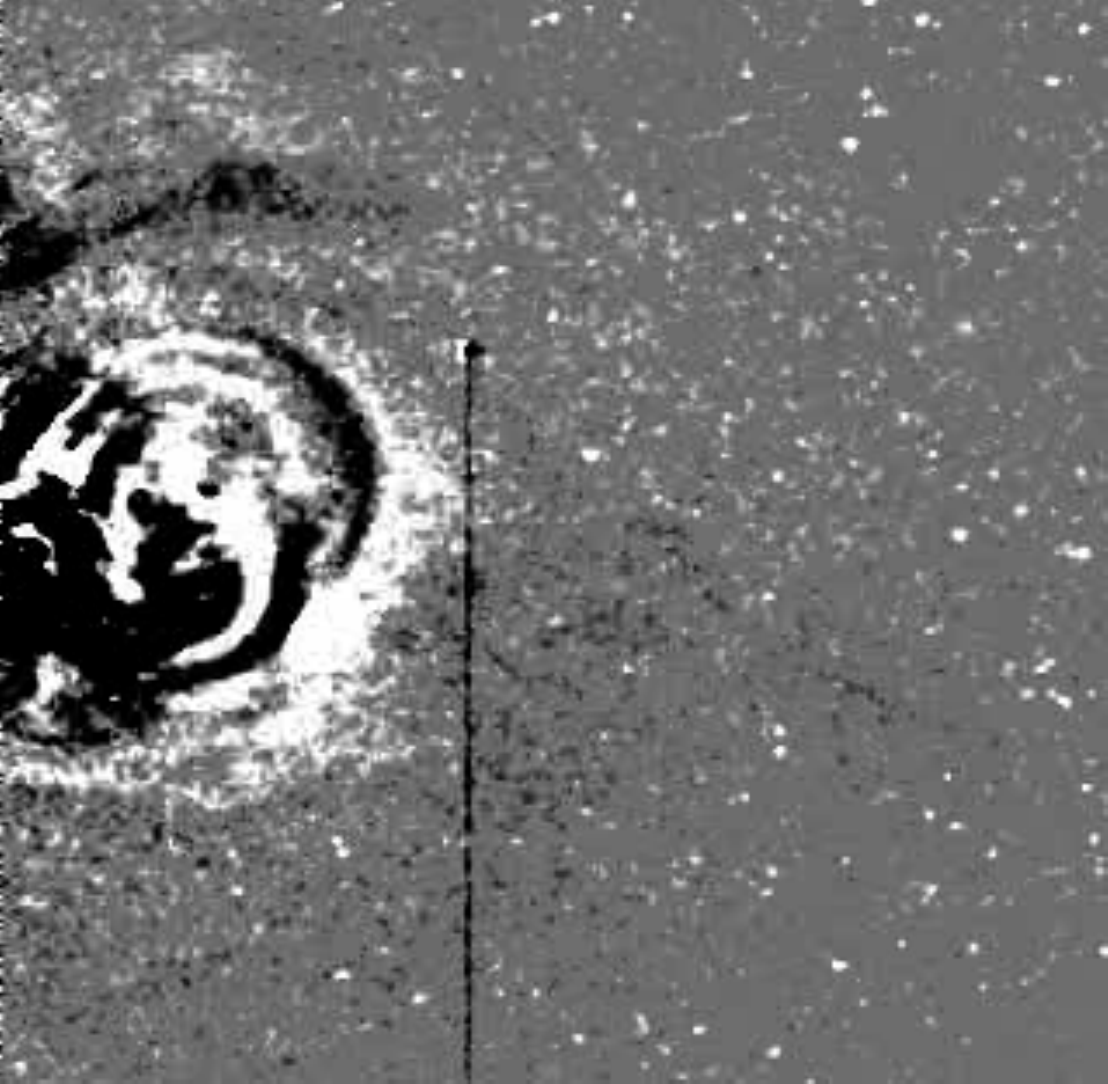}}
\caption{STEREO remote-sensing observations of the 4 December 2009 CME: EUVI-A-304~\AA~at 05:56:15 UT, COR-1B at 09:26 UT with EUVI-B-304~\AA ~at 09:17:04 UT inlaid, COR2-B at 17:09 U and HI1-B at 06:49 UT on 5 December, from top left to bottom right. In the EUVI-A image, the two ends of the filament are indicated with the black arrows.}
\end{center}
\end{figure*}
%%%%%%%%%%%%%%%%%%%%%%%%%

%average arrival time error is 6 for HM and 10.6 for FPhi

\section{Case Study: 4 December 2009 CME Event} \label{sec:Dec2009}

In this section, we give an overview of the analysis of these 20 CMEs by studying in detail one CME observed remotely more than 30$^\circ$ behind the limb. We compare the two analysis techniques and discuss the implications of the CME detection and tracking for CME geometry and the Thomson scattering. 

\subsection{Observations and Fitting}
In early December 2009, STEREO-A and B were separated by almost 130$^\circ$ ($\beta_\mathrm{STA}=63.6^\circ$,  $\beta_\mathrm{STB}=65.8^\circ$) with STEREO-A at 0.97~AU and STEREO-B at 1.06~AU. There was a filament visible in the SECCHI/Extreme Ultraviolet Imager (EUVI) 304~\AA~behind the limb of STEREO-B and disk centered as seen by STEREO-A. It started to rise at around 02:00UT on 4 December and erupted starting at around 06:45UT. It was first visible in COR-1B at 06:50UT and in SOHO/C2 at 08:50 and appeared as a weak halo in COR-1A. It entered into COR-2B field-of-view at 11:55, C3 at 12:20 and HI-1B at 20:49. We show a J-map of this CME in the top panel of Figure~1. Using this J-map, it is possible to track the CME for a total of 85 hours and up to an elongation angle of about 33$^\circ$ (26 datapoints, shown with red circles in the top panel of Figure~1). While erupting from S50, the filament and associated CME were significantly deflected in the low corona towards the equator and the CME enters COR-2B field-of-view propagating at a position angle close to 260--265$^\circ$. As shown in Figure~2, it appears to propagate close to PA 270$^\circ$ well into HI-1 field-of-view. 

We fit the data from STEREO-B/SECCHI with the two fitting methods. The F$\Phi$ fitting yields a best-fit speed of 338 $\pm$ 63 km~s$^{-1}$  and a direction of $30.2^\circ \pm 14.5^\circ$ with respect to the Sun-Earth line. The HM fitting of the same data yields a best-fit speed for the nose of 401 $\pm$ 48.5 km~s$^{-1}$  and a direction of $62.2^\circ \pm 14^\circ$. The derived CME directions, with respect to the observing spacecraft (STEREO-B), are 96$^\circ$ (F$\Phi$) and $128^\circ$ (HM). The middle panel of Figure~1 shows the measured datapoints and the best-fit time-elongation track from the two methods. It can be seen that, although both methods fit the data equally well, there is a difference of more than 30$^\circ$ in the predicted direction of propagation of the CME. As expected from previous theoretical and statistical analyses \cite{Lugaz:2010c,Davies:2012}, for this CME observed behind the limb by STEREO-B, the HM fitting gives a larger angle than the F$\Phi$ fitting method.

The predicted direction based on the HM method is almost directly towards STEREO-A, consistent with a predicted direct hit at the spacecraft. The predicted arrival speed is 401 $\pm$ 48.5 km~s$^{-1}$ with a predicted arrival time of 18:20 UT on 8 December. Based on the F$\Phi$ fitting method, the CME propagates more than 30$^\circ$ off STEREO-A (towards Earth), consistent with a predicted miss. If the CME were to hit STEREO-A, the predicted speed would be 338 $\pm$ 63 km~s$^{-1}$ and the predicted arrival time 12:30 UT on 9 December. As indicated in Figure~3, a fast forward shock was detected {\it in situ} by STEREO-A at 23:38 UT on 8 December. A magnetic cloud was detected from 9 December at 09:00 UT to 10 December at 23:13 UT. The speed of the dense sheath behind the shock was about 330-340~km~s$^{-1}$. The detection of a shock and a magnetic cloud appears to validate the HM fitting which predicts a direct hit at STEREO-A. In addition, the error in the arrival time is slightly better for the HM fitting method as compared to the F$\Phi$ fitting method (about 5 hours for the HM fit versus 13 hours for the F$\Phi$ fit) but the velocity is better predicted by the F$\Phi$ fitting method (see Figure~3). However, for this CME the F$\Phi$ method predicts a miss or, at best, a glancing blow. In fact, it is not consistent to assume that the position is given by the F$\Phi$ approximation and that part of CME 30$^\circ$ away from the CME nose arrives at the same time and with the same speed at 1~AU as the nose of the CME. Therefore, the arrival time and arrival speed predictions derived using the F$\Phi$ approximation should be considered with extreme caution for this case study. 

Additionally, as indicated in Figure~3, there appear to be two distinct magnetic clouds (highlighted in blue and noted as MC1 and MC2) measured {\it in situ}. The first magnetic cloud, with a low inclination ($\sim 10^\circ$ with respect to the solar equatorial plane) is the better candidate to correspond to the low-inclination filament observed in EUVI. We have not been able to find a clear heliospheric (or solar) source of the second, higher inclination ($\sim 60^\circ$) magnetic cloud. While the presence of two clouds has no direct influence on our study which focuses on the arrival time and speed of the density structure ahead of the cloud(s), this example illustrates that many events are not isolated CMEs.

%%%%%Figure 3%%%%%%%%%%%%%%%%%%%
\begin{figure*}[t*]
\begin{center}
{\includegraphics*[height = 12 cm]{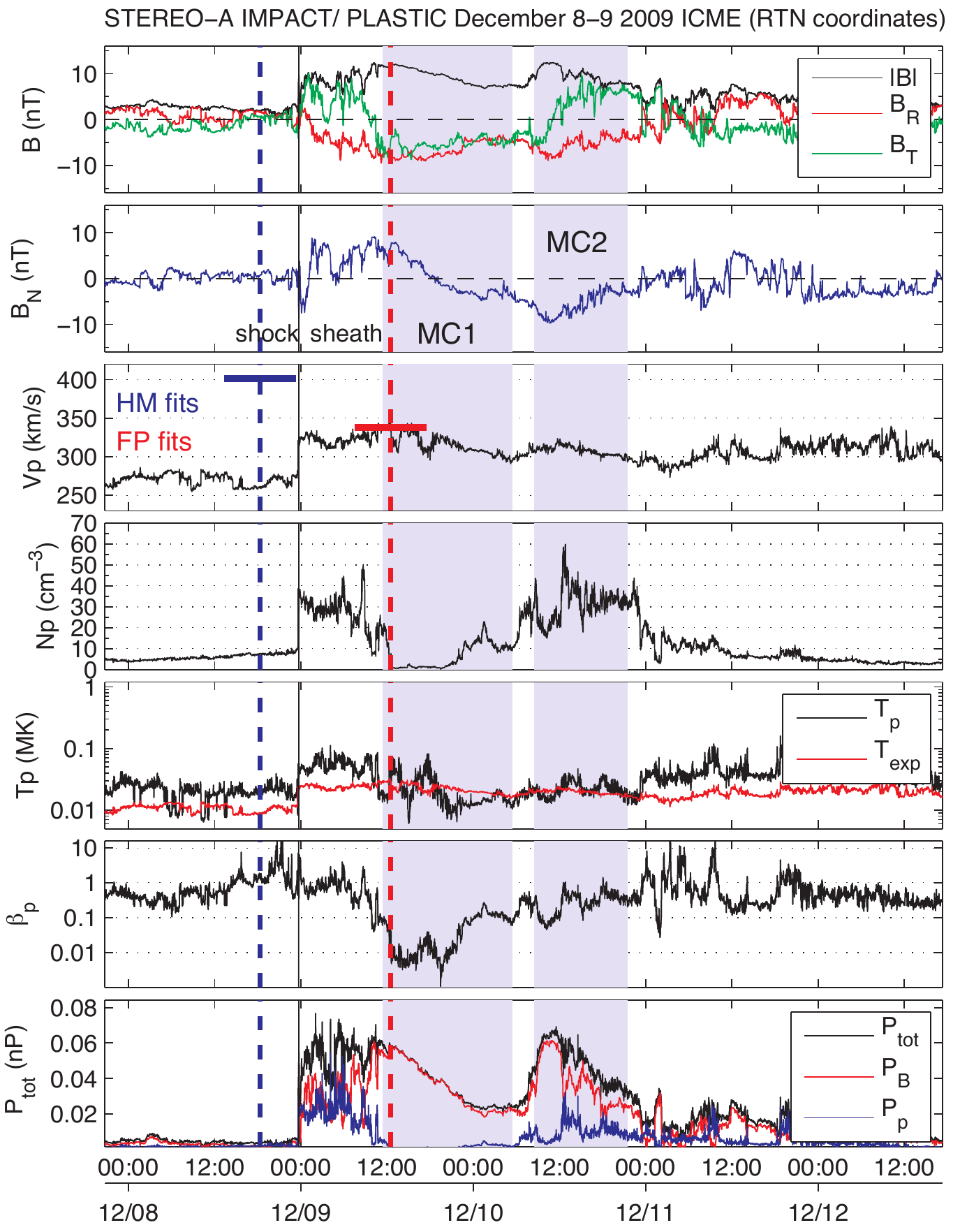}}
\caption{{\it In situ} measurements by STEREO-A IMPACT and PLASTIC of the ICME on 9 December 2009 showing the magnetic field (top 2 panels), proton velocity, density and temperature, the plasma beta and total pressure. The two vertical dashed lines show the predicted arrival time of the front using the HM (blue) and F$\Phi$ fitting (red) methods. The two solid lines show the expected speed and typical error bars of $\pm$ 5 hours for the arrival time for the two methods. MC1 and MC2 are the two magnetic clouds measured at 1~AU, see text for details.}
\end{center}
\end{figure*}
%%%%%%%%%%%%%%%%%%%%%%%%%

\subsection{Further Analysis and Consequences of the Observations for CME Geometry}

While the direction of propagation of the CME seems to be well reproduced with the HM fitting, the predicted arrival speed and arrival time are off. In this respect, it is interesting to investigate in more detail the kinematics of the eruption. To do so, we follow \inlinecite{Wood:2009b}, \inlinecite{Temmer:2011}  and \inlinecite{Rollett:2012} by relaxing the assumption of constant speed. We still assume that the CME propagates radially outward with the direction given by the HM fitting method (62$^\circ \pm 14^\circ$). While deflection is frequent in the corona, prior studies have shown that the assumption of radial propagation holds in the heliosphere \cite{Lugaz:2010b}. In addition, we take an {\it a posteriori} approach similar to that of \inlinecite{Rollett:2012} to constrain the CME parameters with knowledge of the actual arrival speed and arrival time of the CME.

We first focus on matching the arrival speed, measured {\it in situ} to be about 330--340~km~s$^{-1}$. In Figure~4, we show the kinematics of the CME assuming a constant direction of propagation but not a constant speed. The three curves correspond to the best-fit direction of 62$^\circ$ (black) and the lower and upper bounds of the 1-sigma interval 48$^\circ$ (red) and 76$^\circ$ (green). In all three cases, a hit at STEREO-A is predicted since the direction of propagation is within $\pm 20^\circ$ of the spacecraft position. 
It can be seen from the top panel of Figure~4 that the measured speed can be reproduced using the time-elongation data for a CME propagating between 48$^\circ$ and $62^\circ$ from the Sun-Earth line.

We then focus on matching the measured shock arrival time by varying the CME direction of propagation. To do so, we assume, after the last data point, a constant CME speed equal to the average speed of the front over the last 24 hours of observations. For the best-fit direction, the arrival time at STEREO-A under these assumptions is expected to be 19:00 UT on 8 December. The exact arrival time is obtained for a direction of $58^\circ$ corresponding to a speed at STEREO-A of 375 km~s$^{-1}$. In the J-map, for better contrast, the ``black front'' (about 6 hours behind the shock) is being tracked (see top panel of Figure~1). Assuming the arrival time corresponds to that of this black front, the arrival time can be matched with a direction of propagation of 54$^\circ$ and the final speed at STEREO-A is 350~km~s$^{-1}$ in good agreement with the speed measured {\it in situ}. Overall, taking an {\it a posteriori} approach similar to \inlinecite{Rollett:2012}, we find that the measured arrival time and arrival speed can be reproduced assuming  (i) a CME geometry given by the HM approximation of \inlinecite{Lugaz:2009c} and (ii) a constant direction of propagation between 50$^\circ$ and 60$^\circ$ with respect to the Sun-Earth line (or about 10$^\circ$ East of STEREO-A).

From this analysis, there is strong evidence that the CME did not deflect much in the longitudinal direction and remained directed towards STEREO-A. Therefore, STEREO-B/SECCHI was able to image to large elongation angles a CME propagating with a direction of 120$^\circ$ with respect to the Sun-spacecraft line. In addition, the time-height data derived using the HM approximation is consistent with the observed arrival time and the the measured speed in the sheath. 

One of the limitations of the HM and F$\Phi$ fitting models is that they neglect the effect of Thomson scattering.  If the elongation angle corresponds to the intersection of the CME front with the Thomson sphere rather than the tangent to the CME front, as assumed by the HM model, it is still possible for a CME to have a shape which would give the same arrival time and the same kinematics. Such a geometry is shown in green in the bottom panel of Figure~4 next to the assumption of the HM method in pink. The main difference between the two assumptions is about the width of the CME. If we assume that what is observed is the intersection of the CME front with the Thomson sphere, the CME front needs to be significantly wider than if we use the HM approximation. In addition, the signal comes from the ``wings'' of the CME in order to still result in a hit at STEREO-A. It is likely that the real geometry of wide CMEs is somewhere between these two descriptions, although it should be noted that the CME density structure needs to be as wide as 160$^\circ$ for it to intersect with the Thomson sphere.

%%%%%Figure 2%%%%%%%%%%%%%%%%%%%
\begin{figure*}[t*]
\begin{center}
{\includegraphics*[width = 8.5cm]{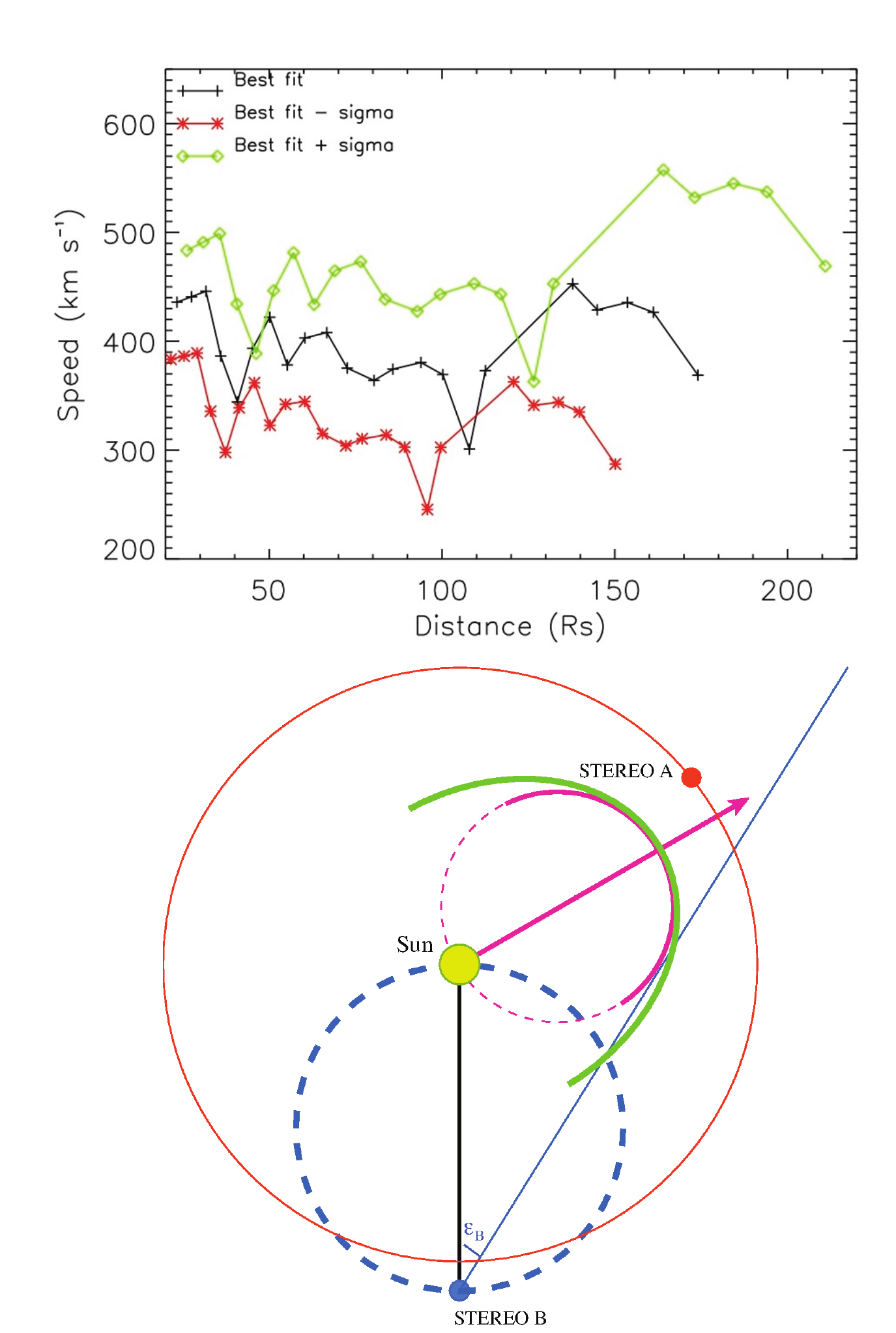}}
\caption{Top: Velocity of the apex of the 4 December 2009 CME assuming that (i) the distance is given by the HM method, (ii) the CME is propagating radially outward at a angle of 48$^\circ$ (red), 62$^\circ$ (black) and 76$^\circ$ (green) from the Sun-Earth line. Assuming precision of 0.2$^\circ$ and 0.5$^\circ$in HI-1 and HI-2, respectively the error bars in the velocity are about $\pm 70-100$~km~s$^{-1}$. Bottom: Sketch of the geometry of the 4 December 2009 CME assuming a viewing angle of 120$^\circ$. The Thomson sphere of STEREO-B is shown in dashed blue. The pink front corresponds to the HM assumption and the green front would give the same kinematics but the signal originates from the intersection of the front with the Thomson sphere.}
\end{center}
\end{figure*}
%%%%%%%%%%%%%%%%%%%%%%%%%

\section{Comparison of the Two Fitting Methods} \label{sec:stat}

The full results of our analysis are shown in Table~2 and top panel of Figure~5 for the CME speed and direction, Table~3 and bottom panels of Figure~5 for the predicted arrival time, hit/miss and the predicted and measured speeds at 1~AU. An online version of the three tables can be found on the electronic supplement to the online article. For the arrival time, we list the shock arrival time, when a shock is present and the start of the ICME when there is no shock. 
Previous studies have shown that the bright structure observed in HI-2 corresponds to the sheath ahead of the magnetic cloud \cite{Wood:2009a,Moestl:2009b,Moestl:2011,Liu:2010b,deForest:2011}. Therefore, for fast CMEs, the shock arrival time, corresponding to the beginning of the sheath region, is the best time to compare to the predicted arrival time from the fitting methods. For slow CMEs which do not drive a shock, we use the starting time of the ICME.
As for the measured speed of the CME, following \inlinecite{Jian:2006}, we report the maximum speed of the ICME, which is usually reached at the start of the ICME (due to the typical decreasing speed profile inside ICMEs) and it usually corresponds to the speed in the sheath. 
The uncertainty in the arrival time due to the uncertainty in the best-fit speed is typically $\pm$ 5--7 hours for the F$\Phi$ method and $\pm$ 8--10 hours for the HM method. However, for most of the CMEs observed in 2009, the uncertainty in the arrival time is $\pm$ 1 day for the following reason. For the CMEs in our studies observed in 2009, when the viewing angle is 90$^\circ$ or more, time-elongation tracks tend to be shorter. This is because a CME propagating behind the limb observed at 1~AU is only at an elongation angle of 35-45$^\circ$, most CMEs in 2009  from our study are only observed up to 35$^\circ$. Previous studies (e.g., see \opencite{Moestl:2011}) have shown that time-elongation tracks must extend well beyond 30$^\circ$ for the uncertainty in the arrival time to be only of the order of $\pm$ 6 hours. 

In the remaining part of the article, we compare the results of the two methods to determine how well they are able to reproduce {\it in situ} measurements. The first criterion under which we judge the fitting methods is whether or not they are able to successfully predict a hit. Following the discussion in section 2.1, we consider that a fitting method predicts a hit if the best-fit propagation angle is within $\pm 20^\circ$ of the Sun-spacecraft line. In a later stage, we consider the predicted arrival time and final speed as compared to {\it in situ} measurements. Predicting these quantities from remote-sensing observations is only important if the fitting methods can correctly identify a CME which is set to hit a spacecraft. In Figure~5, the data points are color-coded with the total spacecraft angular separation from purple for an angle of 45$^\circ$ (Jan. 2008) to dark red for an angle of 140$^\circ$ (Apr. 2010). Square symbols represent events for which both fitting methods predict a hit, triangle symbols correspond to events for which the F$\Phi$ fit predicts a hit but the HM fit predicts a miss and circular symbols to the opposite case. Below, we discuss the different cases.

\begin{table}[htb]
\begin{tabular}{cccccc}
\hline
CME &  F$\Phi$ Speed & F$\Phi$ Direction  & HM Speed & HM Direction &  {\it In situ} Separation \\
 & (km~s$^{-1}$) & ($^\circ$) & (km~s$^{-1}$) & ($^\circ$) & ($^\circ$)\\
\hline
1 & 304 $\pm$ 13.5  & $-28.2  \pm 14.5 $ &   316 $\pm$ 17.5  & $-30.2  \pm 25.5 $ & $-23.6 $\\
2 & 544 $\pm$ 44  & $-33.2  \pm 18 $ &   548~$\pm$ 30.5  & $-33.2  \pm 30.5 $ & $-25.8 $\\
3 & 368 $\pm$ 17  & $34.7  \pm 12.5 $ &   373~$\pm$ 27  & $36.7  \pm 27 $ & 24.3$ $\\
4 & 371 $\pm$ 13   & $-23.8  \pm 5.5 $ &   379~$\pm$ 27.5  & $-33.8  \pm 15 $ & $-29.2 $\\
5 & 314 $\pm$ 5.5   & $11.8  \pm 6.5 $ &   330~$\pm$ 13  & $-0.2  \pm 17 $ & $27.2 $\\
6 & 325 $\pm$ 17   & $-24.8  \pm 11 $ &   329~$\pm$ 18.5  & $-30.8  \pm 16 $ & $-36.2 $\\
7 & 350 $\pm$ 8.5   & $33.7  \pm 10.5 $ &   353~$\pm$ 14.5  & $36.7  \pm 27 $ & 33.3$ $\\
8 & 358 $\pm$ 40   & $38.9  \pm 17.5 $ &   369~$\pm$ 78  & $52.9  \pm 37 $ & 40.1$ $\\
9 & 411 $\pm$ 111   & $32.8  \pm 26.5 $ &   436~$\pm$ 168  & $49.8  \pm 49.5 $ & 43.2$ $\\
10 & 425 $\pm$ 56.5   & $-11.4  \pm 17.5 $ &   431~$\pm$ 37.5  & $-19.4  \pm 16.5 $ & $-42.6 $\\
11 & 351 $\pm$ 62.5  & $-33.2  \pm 17 $ &   379~$\pm$ 35  & $-55.2  \pm 35 $ & $46.6 $\\
12 & 376 $\pm$ 12.5  & $20.9  \pm 6.5 $ &   386 $\pm$ 41  & $33.9  \pm 20.5 $ & $43 $\\
13 & 286 $\pm$ 19  & $-46  \pm 8 $ &   323~$\pm$ 58  & $-74  \pm 12.5 $ & $-55.2 $\\
14 & 407 $\pm$ 114   & $-39.4  \pm 17.5 $ &   505~$\pm$ 174  & $-75.4  \pm 27 $ & $-50.1 $\\
15 & 454 $\pm$ 205   & $-64.6  \pm 27 $ &   699~$\pm$ 201  & $-113.6  \pm 109 $ & $-57 $\\
16 & 283 $\pm$ 138   & $-56.6  \pm 21 $ &   455~$\pm$ 198  & $-107.6  \pm 32.5 $ & $-64.4 $\\
17 & 339 $\pm$ 61   & $32.3  \pm 15 $ &   390~$\pm$ 66  & $61.3  \pm 20.5 $ & 63.1$ $\\
18 & 269 $\pm$ 26   & $-24.5  \pm 52 $ &   306~$\pm$ 66.5  & $-52.5  \pm 20.5 $ & $-64.4 $\\
19 & 338 $\pm$ 63   & $30.2  \pm 14.5 $ &   401~$\pm$ 48.5  & $62.2  \pm 14 $ & $63.7 $\\
20 & 358 $\pm$ 59.5   & $24.3  \pm 11 $ &   439~$\pm$ 137  & $59.3  \pm 22.5 $ & $68.9 $\\
\hline
\end{tabular}
\caption{Best-fit propagation speed and direction with respect to the Sun-Earth Line of the 20 CMEs and separation angle of the spacecraft hit by the CMEs. The separation angle is the one between the impacted spacecraft and Earth and it is therefore different from the one listed in Table~1 between STEREO-A and STEREO-B.}
\end{table}

\begin{table}[htb]
\begin{tabular}{c|ccc|ccc|ccc}
\hline
CME & H/M & F$\Phi$ t$_\mathrm{arr}$ & F$\Phi$ V & H/M  & HM  t$_\mathrm{arr}$ & HM V & t$_\mathrm{arr}$ & V & STEREO\\
\hline 
{\bf 2008}\\
1 & H & 6 Feb 15:50 & 304 & H & 6 Feb 14:50 & 314 & 5 Feb 20:30 & 385 & B\\
2 & H & 29 Apr  19:30 & 544 & H & 29 Apr 20:25 & 541 & 29 Apr 14:10 & 490 & B\\
3 & H & 11 May  11:25 & 368 &H & 11 May 12:30 & 367 & 11 May 06:30 & 350 & A\\
4 & H & 7 Jun  01:30 & 371 &H & 7 Jun  01:30 & 375 & 6 Jun 15:40 & 430 & B\\
5 & H & 5 Jul  11:20 & 314 &{\bf M} & 5 Jul 18:30 & 282 & 5 Jul 00:50 & 360 & A\\
6 & H & 16 Aug 05:50 & 325 & H & 16 Aug 04:50 & 329 & 15 Aug 12:00 & 365 & B\\
7 & H & 4 Sep  22:00 & 350 &H & 4 Sep 21:30 & 353 & 4 Sep 13:20 & 385 & A\\
8 & H & 31 Oct  16:00 & 358 & H & 31 Oct 15:20 & 361 & 31 Oct 12:20 & 400 & A\\
9 & H & 28 Nov  16:50 & 411 & H & 28 Nov 12:30 & 432 & 28 Nov 21:50 & 380 & A\\
10 & {\bf H*} & 31 Dec 10:20 & 425 & {\bf H*} & 31 Dec 21:30 & 386 & 31 Dec 02:00 & 460 & B\\
\hline
{\bf 2009}\\
11 & H & 14 Jan 01:40 & 351 &H & 13 Jan  18:40 & 375 & 13 Jan 5:20 & 400 & B \\
12 & H & 26 Jan 07:50 & 376 & H & 26 Jan 07:10 & 381 & 25 Jan 18:20 & 400 & A\\
13 & H & 17 Jul 06:10 & 286 & H & 16 Jul  20:30 & 306 & 16 Jul 17:10 & 330 & B\\
14 & H & 31 Jul 00:00 & 407 &{\bf M} & 30 Jul 12:40 & 456 & 31 Jul  02:20 & 460 & B\\
15 & H & 1 Oct  09:40 & 454 & {\bf M} & 2 Oct  03:30 & 385 & 2 Oct  17:20 & 360 & B\\
16 & H & 12 Nov 01:50 & 283 & {\bf M} & 12 Nov 01:50 & 332 & 10 Nov  18:50 & 370 & B\\
17 & {\bf M} & 14 Nov 10:05 & 339 & H & 13 Nov 19:00 & 390 & 14 Nov  08:00 & 340 & A\\
18 & {\bf M} & 28 Nov 18:10 & 269 & H & 28 Nov 02:10 & 345 & 27 Nov  12:40 & 400& B\\  
19 & {\bf M} & 9 Dec 12:30 & 338 & H & 8 Dec 18:20 & 401 & 8 Dec  23:40 & 350 & A\\
\hline
{\bf 2010}\\
20 & {\bf M} & 24 Apr 7:30 & 358 & H & 23 Apr 13:00 & 433 & 23 Apr 00:35 & 450 & A\\
\hline
\end{tabular}
\caption{CME number, predicted hit (H) or miss (M), predicted arrival times (UT) and arrival speeds (km~s$^{-1}$) for the F$\Phi$ and HM methods for columns 1 to 7, and actual arrival time and arrival speed and impacted spacecraft for columns 8 to 10. For CME \# 10, both methods predict a direction between STEREO-B and ACE at Earth and the CME impacted both spacecraft, which we note as {\bf H*} (see text for details). The arrival times (measured and predicted) are rounded to the nearest ten minutes.}
\end{table}

%%%%%Figure 2%%%%%%%%%%%%%%%%%%%
\begin{figure*}[t*]
\begin{center}
{\includegraphics*[width=8.cm]{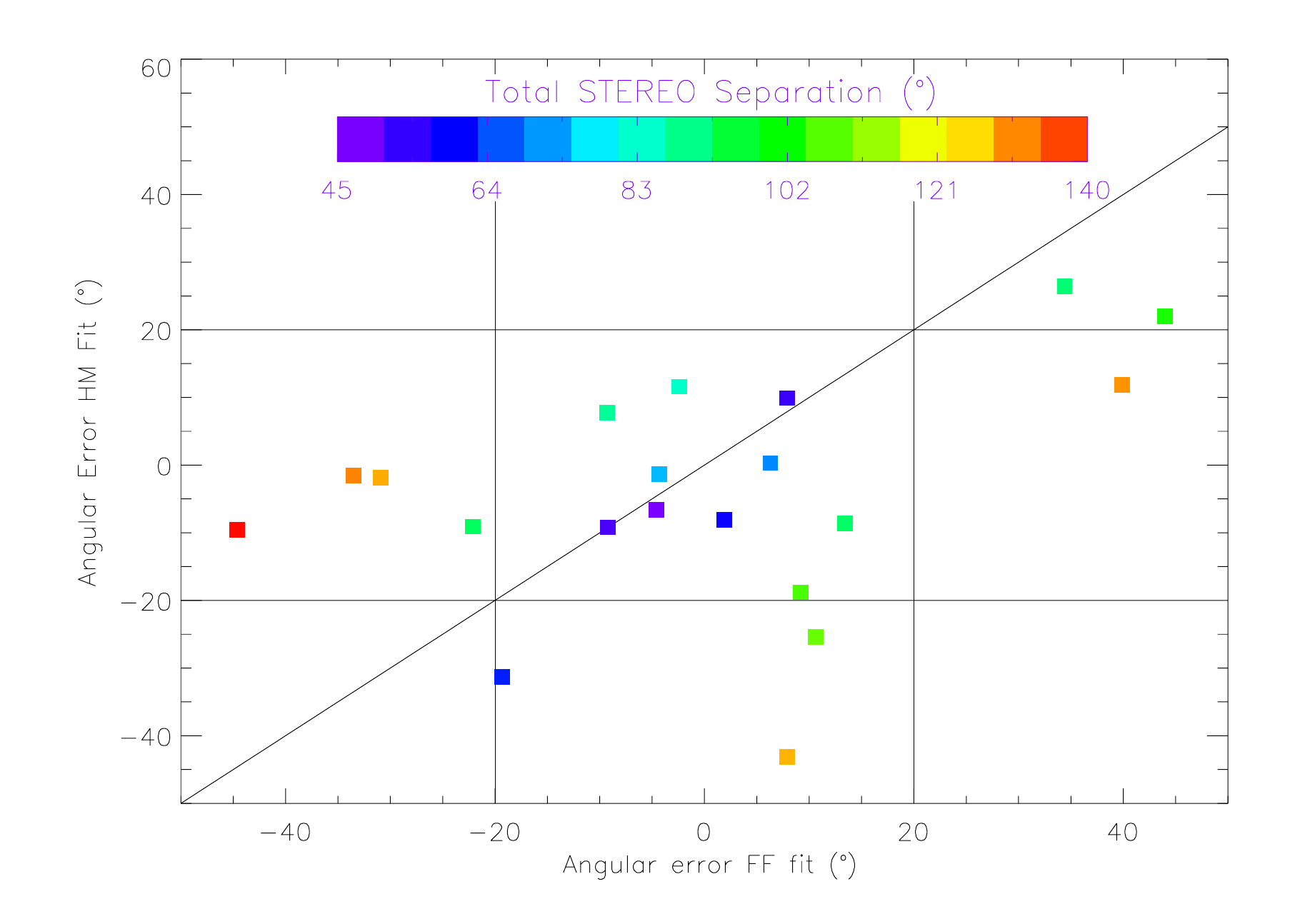}}\\
{\includegraphics*[width=6.cm]{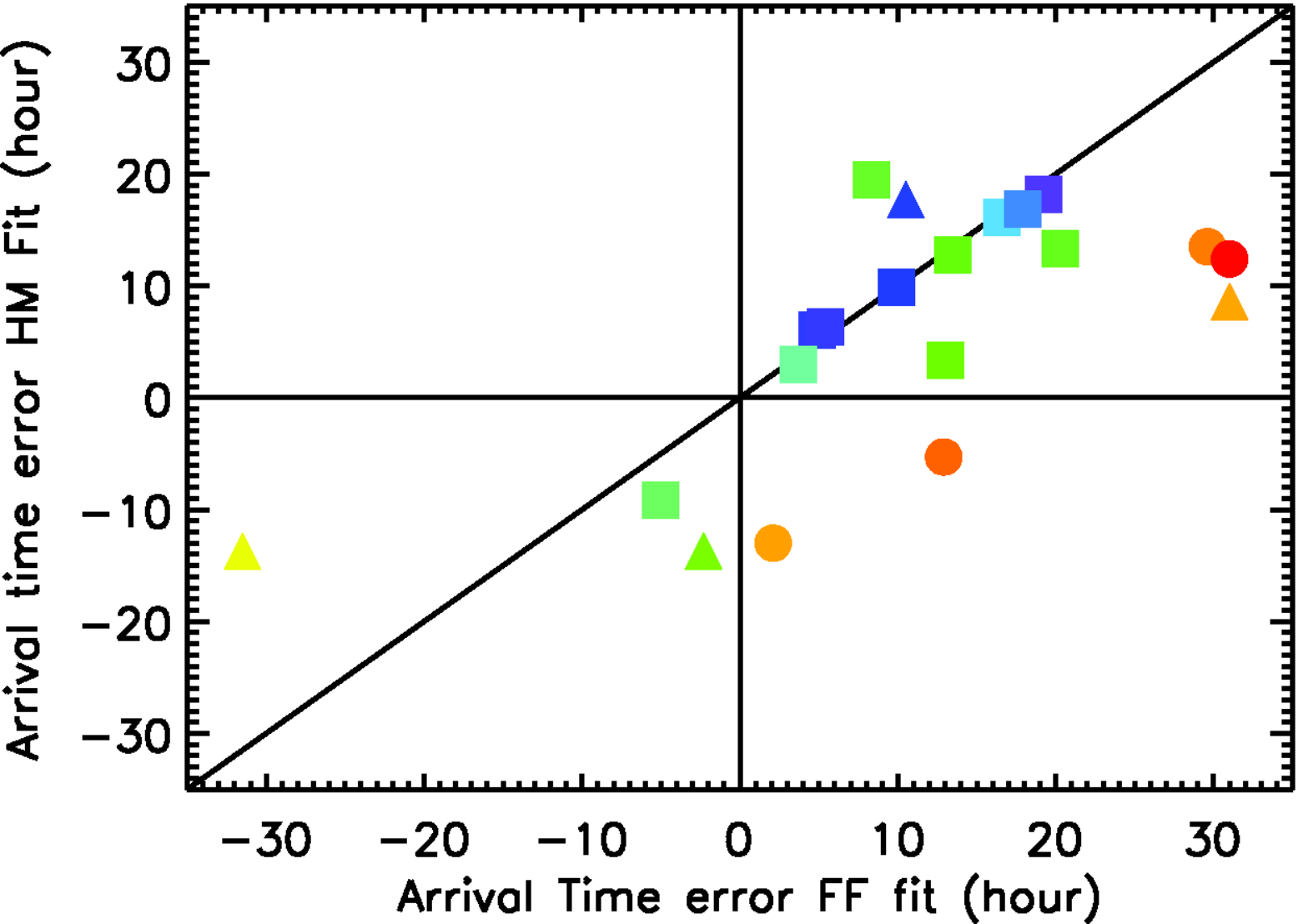}}
{\includegraphics*[width=6.cm]{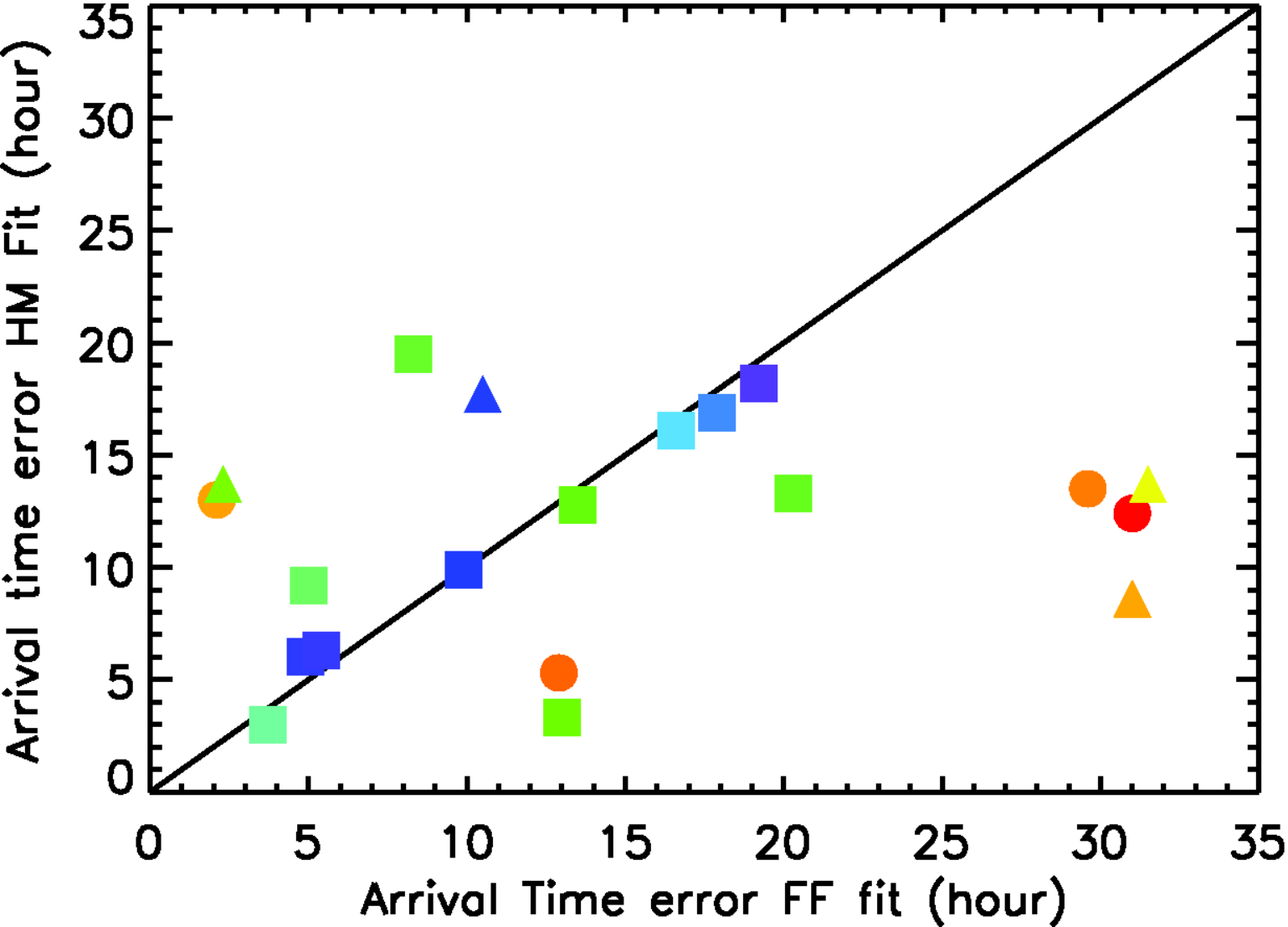}}

\caption{Top: Predicted propagation angle with respect of the hit spacecraft ($x$-axis: F$\Phi$ fitting method, $y$-axis: HM fitting method). Bottom left: Same as top but for the error in the arrival time. Bottom right: Same as left but for the absolute value of the error in arrival time. In all panels, the symbols are color-coded with the total spacecraft separation (purple: 45$^\circ$, dark red: 140$^\circ$). In the bottom panels, square symbols are for CME events for which both methods would have predicted a hit, triangle those for which only the F$\Phi$ fitting method would have predicted a hit and circle those for which only the HM fitting method would have predicted a hit.}
\end{center}
\end{figure*}
%%%%%%%%%%%%%%%%%%%%%%%%%

\subsection{Events Successfully Predicted by Both Methods}

Nine out of the ten events in 2008 are successfully predicted by both methods (the exception is CME 5) but the same is true for only three out of the ten events in 2009--2010 (first 3 events). 
For nine out of these 12 events, the arrival time predicted from the two fitting methods is nearly identical (see square symbols in the bottom panels of Figure~5). The exceptions are events 10, 11 and 13 (all for separation greater than 85$^\circ$). This finding is also confirmed by Earth-directed events, for which both methods have essentially given the same predictions up until the beginning of 2011. The arrival time of events 11 and 13 is significantly better reproduced with the HM fitting method than with the F$\Phi$ method. This comes from the fact that the HM best-fit velocity is usually higher than the best-fit velocity from the F$\Phi$ method (see also \opencite{Davies:2012}). It is also consistent with the fact that for viewing angles greater than 80$^\circ$, the HM fitting method is expected to give better results than the F$\Phi$ method \cite{Lugaz:2010c}. Event 10 is different because both methods predict a CME direction more than 25$^\circ$ away from the spacecraft (towards Earth), i.e. both methods predict a miss. However, this event was both observed {\it in situ} by ACE and STEREO-B, so we consider that the best-fit angle gives an accurate prediction. 

As noted before, the back of the white front in the J-map (``black front'') is being tracked. We consider that this front is 4--6 hours behind the CME front as seen in J-maps. Therefore, there is a bias of the two methods towards finding later arrival times. In fact, if we look at the average (signed) arrival time error, it is +6 hours for the HM and +10.6 hours for F$\Phi$ fitting methods. Only five events are predicted early by one or both methods. We believe part of this average error in the arrival time is directly attributable to the fact that we tracked the ``black front'' in the J-map. 

\subsection{Hits  Predicted by the F$\Phi$ but not HM Fits}

For CMEs 5, 14, 15 and 16, the best-fit CME direction from the F$\Phi$ fit is within 20$^\circ$ of the STEREO spacecraft but the best-fit direction from the HM is more than 20$^\circ$ away. 
CME 5 is a very faint event for which the best-fit angle from both methods indicate an expected hit at L1, which was not recorded. It is likely that the assumption of wide CME used in the HM fitting method does not apply for this CME. CME 14 is a relatively fast event (460 km~s$^{-1}$ at 1 AU from the {\it in situ} measurements) observed at a viewing angle of about 100$^\circ$. For CMEs faster than the solar wind, the assumption of constant propagation speed used in the fitting methods results in an additional error: the physical deceleration due to the interaction with the solar wind \cite{Cargill:2000,Vrsnak:2010,Davis:2010} is fitted as a geometrical deceleration associated with a large viewing angle. This effect is particularly pronounced for the HM fitting method \cite{Kintner:2011} because it has, intrinsically, larger errors than the F$\Phi$ fitting method.

For CMEs 15 and 16, the HM fits give a better match for the arrival time than the F$\Phi$ fits. Event 15 corresponds to a small track due to a data gap starting at around 22:00 UT on 28 September (when the CME is around 15$^\circ$ elongation). Previous works have shown that density transients must be imaged to much larger elongations to be fitted with accuracy \cite{Williams:2009,Moestl:2011}. The filament eruption originated from the eastern side of the solar disk as seen from STEREO-B, corresponding to an angle of propagation likely to be greater than 130$^\circ$. Event 16 is another filament eruption from behind the eastern limb as seen in STEREO-A, but it is unclear from STEREO-B images where it originated from. In both events, the HM method is better mostly due to the correction in the arrival time of \inlinecite{Moestl:2011}. However, the best-fit angles obtained by the HM fitting method are unrealistically large ($\sim 170^\circ$ viewing angle). The fact that HM fitting method yields a better arrival time estimate than the F$\Phi$ fitting method is probably just coincidental.

\subsection{Hits  Predicted by the HM but not F$\Phi$ Fits}

For the last four events of our list (17, 18, 19 and 20), a hit was only predicted with the HM fit. For the last three events (18, 19, 20), the arrival time is best predicted with the HM fitting method as well, whereas for event 17, the error is smaller using the F$\Phi$ fit. However, for event 17, the CME direction of propagation predicted by the F$\Phi$ fitting is more than 30$^\circ$ away from the spacecraft, which makes it unlikely that the F$\Phi$ fitting predicts a hit. The filament eruption for event 17 is seen almost as disk-centered from STEREO-A, which corresponds  to the direction predicted by the HM fit. The larger error originates from the larger speed as compared with the F$\Phi$ fit. 

\section{Discussion and Conclusions} \label{sec:conclusions} 

In this article, we have examined STEREO-impacting CMEs, combined {\it in situ} measurements with remote observations by STEREO/SECCHI and compared the prediction for hit/miss, arrival time and speed for the two most commonly used fitting methods. We have started from a list of 47 ICMEs observed using PLASTIC and IMPACT from January 2008 to June 2010, during which time the separation between the two STEREO spacecraft increased from 45$^\circ$ to 140$^\circ$. We have been able to identify the remote-sensing observations for 20 ICMEs ($\sim 40 \%$). For some ICMEs for which we were unable to identify a counterpart in the remote-sensing observation, this was due to data gaps or poor observations leading to short tracks.
It is also likely that some of the ICMEs identified at 1~AU were small flux ropes embedded in the solar wind or in stream interaction regions (SIRs). This type of ICMEs is unlikely to be observed behind the limb.

The proportion of ICMEs for which we find remote-sensing observations decreases with increasing viewing angle from 10/14 in 2008 (separation less than 90$^\circ$) to only 1/12 in 2010 (separation greater than 130$^\circ$). However, we find six CMEs viewed more than 20$^\circ$ behind the limb, which can be tracked into HI-2 field-of-view. Five of these six CMEs were filament eruptions, hence, associated with wide CMEs. Therefore, we predict that some Earth-directed CMEs will be successfully tracked by SECCHI until early 2013, when the STEREO--Earth separation will be comparable to the STEREO-A--STEREO-B separation in early 2010. It should be noted that the first CME observed by SECCHI propagated 20--30$^\circ$ behind the limb \cite{Lugaz:2009b} and was imaged to 40$^\circ$ elongation angle. If the measured signal in the HI instruments come from the part of the CME closest to the Thomson sphere, only very wide CMEs might be observed up to these large viewing angles (similar to the original argument of \opencite{Vourlidas:2006}). If the Thomson scattering effects are relatively negligible due to the extended and non-uniform density structure of the CME front, then the detected signal is likely to originate from the tangent to the CME front as proposed by \inlinecite{THoward:2009b} and \inlinecite{Lugaz:2009c}. In this case, less wide ($\sim 60^\circ$) Earth-directed CMEs might be observed in 2012.

We have compared the fitting method based on the Fixed-$\Phi$ approximation \cite{Sheeley:1999}, where the CME apex is assumed to be observed at all times, with that based on the Harmonic Mean approximation \cite{Lugaz:2009c}, where the tangent to a circular CME front is assumed to be observed.
We confirm that both methods give nearly identical results for the propagation angle of CMEs viewed less than 80$^\circ$ away from the observing spacecraft. We have also found that this holds true for the predicted arrival time and final speed of CMEs. We show evidence, based on 20 ICMES, that these methods provide direction and speed consistent with {\it in situ} measurements. For viewing angles larger than $90^\circ$, we find significant differences between both methods. We find some anecdotal evidence that  the HM fitting method might be better adapted to very large viewing angles as compared to the F$\Phi$ fitting method: the arrival time predicted using the HM fit is better (4/8) or identical (2/8) than the arrival time predicted from the F$\Phi$ fit in six out of eight CMEs observed behind the limb. However, for some fast CMEs, the HM fitting method gives unrealistic results for the CME speed and direction, for the following reason. The constant velocity assumption of these methods causes CME deceleration (acceleration) to be interpreted as geometrical deceleration (acceleration), resulting in errors in the best-fit direction of propagation. This effect has been found to be more pronounced for the HM fitting method as compared to the F$\Phi$ method \cite{Kintner:2011}. These results should be confirmed in further studies using Earth-directed CMEs as well as CMEs which impacted planetary missions such as those dedicated to Mercury, Venus and Mars. In addition, further studies should look into more details at the {\it in situ} characteristics (speed, duration, strength, orientation, etc...) of ICMEs which were identified in remote-sensing observations as compared to those for which we could not find a corresponding track in J-maps.

 \begin{acks}
The authors would like to thank the anonymous referee for useful comments and suggestions, which helped improve the clarity of the manuscript. 
N.~L. was supported during this work by NSF grants AGS-0819653, AGS-1239699 and AGS-1239704 and NASA grants NNX-07AC13G, NNX-08AQ16G and NNX-12AB28G. P.~K. performed research for this work under a Research Experience for Undergraduates (REU) position at the University of Hawaii's Institute for Astronomy and funded by NSF. L.K.J.'s work was funded by NASA's SMD as part of the STEREO project, including the IMPACT investigation. This work was also partially supported by NASA STEREO program through grant NAS5-03131 to UC Berkeley and UNH and has received funding from the European Union Seventh Framework Programme (FP7/2007-2013) under grant agreement 263252 [COMESEP]. C.~M. was supported by a Marie Curie International Outgoing Fellowship within the 7th European Community Framework Programme (PIOF-GA-2010-272768 [WILISCME]).
SOHO and STEREO are projects of international cooperation between ESA and NASA. 
 The SECCHI data are produced by an international consortium of Naval Research Laboratory, Lockheed
  Martin Solar and Astrophysics Lab, and NASA Goddard Space Flight
  Center (USA), Rutherford Appleton Laboratory, and University of
  Birmingham (UK), Max-Planck-Institut f{\"u}r Sonnensystemforschung
  (Germany), Centre Spatiale de Liege (Belgium), Institut d'Optique
  Th{\'e}orique et Appliqu{\'e}e, and Institut d'Astrophysique
  Spatiale (France).  
\end{acks}

\bibliographystyle{spr-mp-sola-cnd}

\end{article}

\end{document}